\documentclass{jfm}
\usepackage{color}
\usepackage{amstext}
\usepackage{amssymb}
\usepackage{graphicx}
\usepackage{amsmath}

\newcommand\const{\mathrm{const}}
\newcommand\Div{\mathrm{div}\,}

\newcommand\vX{\boldsymbol{X}}
\newcommand\vP{\boldsymbol{P}}

\newcommand\vV{\boldsymbol{V}}
\newcommand\va{\boldsymbol{a}}
\newcommand\vb{\boldsymbol{b}}

\newcommand\vf{\boldsymbol{f}}
\newcommand\vh{\boldsymbol{h}}

\newcommand\vK{\boldsymbol{K}}

\newcommand\vu{\boldsymbol{u}}
\newcommand\vv{\boldsymbol{v}}

\newcommand\vx{\boldsymbol{x}}

\newcommand\vp{\boldsymbol{p}}
\newcommand\vq{\boldsymbol{q}}
\newcommand\vQ{\boldsymbol{Q}}
\newcommand\vg{\boldsymbol{g}}

\newcommand\vxi{\boldsymbol{\xi}}

\newcommand\vkappa{\boldsymbol{\kappa}}

\renewcommand{\theequation}{\arabic{section}.\arabic{equation}}

\begin{document}

 {\title[Two-timing Hypothesis, Distinguished Limits, Drifts, and Vibrodiffusion] {Two-timing Hypothesis, Distinguished Limits, Drifts, and Vibrodiffusion for Oscillating Flows}}

\author[V. A. Vladimirov]{V.\ns A.\ns V\ls l\ls a\ls d\ls i\ls m\ls i\ls r\ls o\ls v}

\affiliation{DOMAS, Sultan Qaboos University, Oman and DAMTP, University of Cambridge, UK}


\setcounter{page}{1}\maketitle \thispagestyle{empty}

\begin{abstract}
In this paper we develop and use the two-timing method for a systematic study of a scalar advection caused by a general oscillating velocity field. Mathematically, we study and classify the multiplicity of distinguished limits and asymptotic solutions produced in the two-timing framework. Our calculations go far beyond the usual ones, performed by the two-timing method.  We do not use any additional assumptions, hence our study can be seen as a test for the validity and sufficiency of the two-timing hypothesis.
Physically, we derive the averaged equations in their maximum generality (and up to high orders in small parameters) and obtain qualitatively new results.
Our results are:
(i) the dimensionless advection equation contains \emph{two independent dimensionless small parameters}: the ratio of two time-scales and the spatial amplitudes of oscillations;
(ii) we identify a sequence of \emph{distinguished limit solutions} which correspond to the successive degenerations of a \emph{drift velocity};
(iii) for a general oscillating velocity field we derive the averaged equations for the first \emph{four distinguished limit solutions};
(iv) we  show, that \emph{each distinguish limit solution} produces an infinite number of \emph{parametric solutions} with a Strouhal number as the only large parameter; those solutions differ from each other by the slow time-scale and the velocity amplitude;
(v) the striking outcome of our calculations is the inevitable appearance of \emph{vibrodiffusion}, which represents a Lie derivative of the averaged tensor of quadratic displacements;
(vi) our main methodological result is the introduction of a logical order/classification of the solutions; we hope that it opens the gate for applications of the same ideas to more complex systems;
(vii) five types of oscillating flows are presented as examples of different drifts and vibrodiffusion.

\end{abstract}

\section{Introduction}

This paper is both mathematical (aimed to develop the two-timing method in the form introduced in \cite{Vladimirov2, Yudovich,
Vladimirov1}), and physical (targeted to obtain the multiple versions of averaged equations for a passive scalar in an oscillating flow). Mathematically, we study a hyperbolic PDE of the first order. The use of the two-timing assumption  leads to a problem with two independent small scaling parameters  representing the ratio of two time-scales and the vibrational amplitude of oscillations.
The asymptotic solutions require the choice of  asymptotic paths in the space of parameters. Usually some `lucky' path is presented for a particular problem, while the only reason for choosing this path is the successful calculation of the main term in asymptotic solution, see \emph{e.g.} \cite{Kevorkian}.
Such `lucky' asymptotic paths are called  \emph{distinguished limits}, while the related solutions represent  \emph{distinguished limit solutions} (DLS's).
At the same time any studies of multiplicity of distinguished paths and interrelations between them are always avoided.
The  different approach is given in \cite{Yudovich, VladimirovDr1}, where \emph{an inspection procedure} for calculations of distinguished paths is proposed and the concepts of `strong',  `moderate', and `weak' oscillations (based on the varying of vibrational velocity amplitude) is introduced and exploited. However, this approach can be seen as `too rigid' and will be further developed in this paper.

Below, we present a more general asymptotic procedure (than in \cite{Yudovich, VladimirovDr1}) based on simultaneous varying of the slow time-scale and vibrational velocity amplitude, which from our point of view is more flexible, self-consistent, and leads to a convenient  classification of solutions. The particularly interesting  problem here is to identify and  classify a large number of similarly expressed \emph{parametric solutions} which correspond to the same distinguished limit, but can be obtained with a different choice of small parameters. Since we do not introduce any additional assumptions (except the two-timing hypothesis itself) our paper can be seen as the test to validate the two-timing method. Any failure to interpret our result physically can be explained by insufficiency of two time scales only.

Our analytic calculations are straightforward by their nature, but they do include a large number of integration by parts and algebraic transformations;
the performing of such calculations in a general formulation represents a `champion-type' result by itself.
These calculations are presented in Appendix to this paper, they are also described in detail (among many other calculations) in the arXiv papers by \cite{VladimirovDr1, VladimirovDr2} quoted below as I and II. As far as we know, the two-timing solutions for high approximations has never been calculated before and the degree of similarity of DLS's corresponding to different distinguished paths has never been studied.

As the first step in our analysis we introduce the inverse Strouhal number $1/St$ as a basic small parameter; the appearance of $St$ means that the scaling is not unique and we chose it to vary a velocity scale.
Then, we introduce an infinite sequence of the distinguished limits and DLS's, which correspond to the successive degeneration of a drift velocity,
while the \emph{order} of this degeneration is chosen to enumerate the distinguished limits.  The calculations of DLS's produce the successive approximations for a drift velocity as well as a qualitatively new `diffusion-like' term, which we call \emph{vibrodiffusion}.
Remarkably, all the coefficients in averaged equations are universal; they are the same in DLS's of different orders and only `moving' to lower approximations with the increasing of the order of  DLS's.

From the physical viewpoint our study can be seen as a broader  general view on the studies of  drift velocities with emphasis on their different appearances in averaged equations; it is complemented by an unexpected but inevitable appearance of vibrodiffusion. For different types of flows vibrodiffusion can correspond to diffusion or `anti-diffusion' or to some intermediate cases. The appearance of vibrodiffusion as a \emph{Lie-derivative} of an averaged \emph{quadratic displacement tensor} represents a major challenge for explanation and analysis.
As the main result in the physical implementation of DLS's  we discover \emph{one-parametric} family of solutions (corresponding to each DLS), which leads to the possibility of considering different time-scales and amplitudes of velocity.
Our results are particularly striking for the case of a given flow with purely periodic velocity oscillations. The question we address is: what is the reason for choosing of any particular slow time-scale? Our answer is:  the slow time-scale is uniquely determined by the magnitude of a given velocity in terms of $St$. For a given order of velocity we obtain the unique slow time-scale. Solutions with different orders of velocity are physically different even if they correspond to the same functional dependence of velocity on spatial coordinates and time.
To illustrate the different appearances of drift and vibrodiffusion  we present five examples of flows.

The drift, in its main approximation, is a classical and well-studied phenomenon, see  \cite{Stokes, Maxwell, Lamb, LH, Darwin, Batchelor, Lighthill, McIntyre, Craik, Grimshaw, Craik0, Hunt1, Eames, Buhler}. In our consideration we just place drift flows into a more general content, which show its different options and appearances.
At the same timethe main purpose of this paper is \emph{the
introduction of a new systematic and general viewpoint on the hierarchy of distinguished limits, drifts, and vibrodiffusion}.
This new viewpoint helps us to organize and unify a number of results, including that of I, II, \cite{Vladimirov2, Yudovich, VladimirovL, VladProc}.


\section{Formulation of Problem and Two-Timing Assumption}

A fluid flow is given by its velocity field $\vv^*(\vx^*,t^*)$, where $\vx^*=(x_1^*,x_2^*,x_3^*)$ and $t^*$ are cartesian coordinates and time, asterisks stand for dimensional variables.
We suppose that this field is sufficiently smooth, but we do not suppose that it satisfies any equations of motion.
The dimensional advection equation for a scalar field $a(\vx^*,t^*)$ is
\begin{eqnarray}
a_{t^*}+(\vv^*\cdot\nabla^*) a=0,\qquad a_{t^*}\equiv \partial a/\partial {t^*}
\label{exact-1}
\end{eqnarray}
This equation describes the motions of
a lagrangian marker in either an incompressible or compressible fluid and the advection of a passive scalar admixture
with concentration $a(\vx^*,t^*)$ in an incompressible fluid.
The hyperbolic equation (\ref{exact-1}) has characteristics curves (trajectories) $\vx^*=\vx^*(t^*)$ described by an ODE
\begin{eqnarray}
{d\vx^*}/{dt^*}=\vv^*,\qquad \vx^* |_{t=0}=\vX^*
\label{exact-1a}
\end{eqnarray}
where $\vx^*$ and $\vX^*$ are eulerian and lagrangian coordinates.
The classical description of  drift motion follows after the integration of (\ref{exact-1a}),
however for higher approximations it requires very bulky operations with lagrangian displacements.
Therefore in this paper we solve the equation (\ref{exact-1}) which allows to use the eulerian average operation
and to simplify calculations. The field $\vv^*$ is oscillating in time and possesses the characteristic scales of velocity $U$, length $L$, and frequency $\omega^*$.
These three parameters give a Strouhal number $St= L\omega^*/U$, hence the dimensionless variables and parameters (written without asterisks) \emph{are not unique}; we use the following set
\begin{eqnarray}
&& \vx^*= L\vx,\ t^*=(L/U)t,\ \omega^*=(U/L)\omega,\ \vv^*=U\vv\label{variables}
\end{eqnarray}
while in $\vv$ has the `two-timing' functional form
\begin{eqnarray}
&& \vv =\omega^\beta\vu(\vx, s, \tau),\  \tau=\omega t,\quad s= t/\omega^\alpha;\quad \vu\sim O(1)\label{variables1}
\end{eqnarray}
with constants
\begin{eqnarray}
&& \alpha=\const >-1,\ \beta=\const<1,\ \omega=St\gg 1 \label{exact-2a}
\end{eqnarray}
The $\tau$-dependence of $\vu$ is always $2\pi$-periodic, while its $s$-dependence is arbitrary.
The values of constants  $\alpha$ and $\beta$ will be defined later in distinguished limits.
The restriction $\alpha>-1$  makes the variable $s$ `slow' in comparison with $\tau$,  while $\beta<1$ gives the smallness of vibrational spatial amplitude; $\omega\equiv St$ is considered as a large parameter.

In dimensionless  variables (after the use of chain rule) (\ref{exact-1}) takes the form
\begin{eqnarray}
&&\omega a_\tau+ \frac{1}{\omega^\alpha} a_s+\omega^\beta({\vu}\cdot\nabla)a= 0, \qquad {\partial}/{\partial{t}}=\omega\partial/\partial\tau+\frac{1}{\omega^\alpha}\partial/\partial s\label{exact-6}
\end{eqnarray}
where the subscripts $\tau$ and $s$  stand for partial derivatives; $s$ and $\tau$ represent two mutually dependent time-variables, which are called \emph{slow time} and \emph{fast time}.
Equation (\ref{exact-6}) can be rewritten in the form containing two independent small parameters
\begin{eqnarray}
&&a_\tau+\varepsilon_2 a_{s}+\varepsilon_1 \vu\cdot\nabla a= 0;\quad \varepsilon_1\equiv \omega^{(\beta-1)}, \quad \varepsilon_2\equiv 1/\omega^{(\alpha+1)}
\label{exact-6a}
\end{eqnarray}
Hence, we operate in the plane $(\varepsilon_1, \varepsilon_1)$, where we study asymptotic limit
\begin{eqnarray}
&&(\varepsilon_1,\varepsilon_2)\to (0,0)
\label{paths}
\end{eqnarray}
Different asymptotic paths $(\varepsilon_1,\varepsilon_2)\to (0,0)$ may produce different solutions, such paths are called \emph{distinguished limits}.

The key suggestion of the two-timing method is to consider $\tau$ and $s$ as mutually independent time variables. As a result, we convert \eqref{exact-6}, \eqref{exact-6a} from a PDE with independent variables $t$ and $\vx$  into a PDE with the extended number of independent variables $\tau, s$ and $\vx$. Then the solutions of \eqref{exact-6}, \eqref{exact-6a} must have a functional form:
\begin{eqnarray}
&& a=a(\vx, s, \tau)\label{exact-3}
\end{eqnarray}
It should be emphasized, that without this suggestion a functional form can be different from \eqref{exact-3}, since the presence of $St$ allows to build an infinite number of different time-scales, not just $\tau$ and $s$.
In order to make further analytic progress, we require some specific notations and agreements.
In this paper we assume that any dimensionless function $f(\vx,s,\tau)$ has the following properties:

(i)  $f\sim {O}(1)$ and  all its $\vx$-, $s$-, and $\tau$-derivatives of required below orders are also ${O}(1)$, the only exception is $\vv\sim O(\omega^\beta)$ in \eqref{variables1};

(ii) $f$  is $2\pi$-periodic in $\tau$, i.e.\ $f(\vx, s, \tau)=f(\vx,s,\tau+2\pi)$ (about this simplification see the Discussion section);

(iii) $f$ has an average given by
$$
\langle {f}\,\rangle \equiv \frac{1}{2\pi}\int_{\tau_0}^{\tau_0+2\pi}
f(\vx, s, \tau)\, d \tau\equiv \overline{f}(\vx,s) \qquad \forall\ \tau_0=\const;
$$

(iv) $f$ can be split into averaged and purely oscillating parts, $f(\vx, s, \tau)=\overline{f}(\vx, s)+\widetilde{f}(\vx, s, \tau)$ where  \emph{tilde-functions} (or  purely oscillating functions) are such that $\langle \widetilde f\, \rangle =0$ and the \emph{bar-functions} $\overline{f}(\vx,s)$ (or averaged functions) are $\tau$-independent;

(v)  we introduce a special notation $\widetilde{f}^{\tau}$ with superscript $\tau$ for  the
\emph{tilde-integration} of tilde-functions, such integration keeps the result in the tilde-class:
\begin{equation}
\widetilde{f}^{\tau}\equiv\int_0^\tau \widetilde{f}(\vx,s,\tau')\, d \tau'
-\frac{1}{2\pi}\int_0^{2\pi}\Bigl(\int_0^\mu
\widetilde{f}(\vx,s,\tau')\, d \tau'\Bigr)\, d \mu.
\label{ti-integr}
\end{equation}

\section{Distinguished Limits and Related Solutions}
It can be shown (see I,II) that there are series of distinguished limits for the equation (\ref{exact-6a}) with independent variables $\tau, s, \vx$, which represent the one-parametric distinguished path $\varepsilon_1=\varepsilon$, $\varepsilon_2=\varepsilon^n$
with an integer $n>0$:
\begin{eqnarray}
&&a_\tau+\varepsilon^n a_{s}+\varepsilon \vu\cdot\nabla a= 0
\label{exact-6c}
\end{eqnarray}
The first four Distinguished Limits DL($n$) with $n=1,2,3,4$ are:

DL(1)  $s= t$;  $\vu=\overline{\vu}(\vx,s)+\widetilde{\vu}(\vx,s,\tau)$, here both `bar' and `tilde' parts of $\vu$ are not zero;

DL(2) $s=\varepsilon t$;\  $\vu=\widetilde{\vu}(\vx,s,\tau)\neq 0$, while $\overline{\vu}(s,\vx)\equiv 0$;

DL(3)  $s=\varepsilon^2 t$;\  $\vu=\widetilde{\vu}(\vx,s,\tau)$,  while $\overline{\vu}(s,\vx)\equiv 0$, and $\overline{\vV}_0\equiv 0$;

DL(4) $s=\varepsilon^3 t$;\  $\vu=\widetilde{\vu}(\vx,s,\tau)$, while $\overline{\vu}(s,\vx)\equiv 0$, $\overline{\vV}_0\equiv 0$, and $\overline{\vV}_1\equiv 0$.

\noindent We introduce notations:
\begin{eqnarray}
&&\vV_0\equiv \frac{1}{2}[\widetilde{\vu},\widetilde{\vxi}],\quad
\overline{\vV}_1\equiv\frac{1}{3}\langle[[\widetilde{\vu},\widetilde{\vxi}],\widetilde{\vxi}]\rangle,\quad \widetilde{\vxi}\equiv\widetilde{\vu}^\tau
\label{4.18}
\end{eqnarray}
where we use the notation for commutator $[\vf,\vg]\equiv(\vg\cdot\nabla)\vf-(\vf\cdot\nabla)\vg$ of any vector-functions $\vf$ and $\vg$. In all cases we are looking for the solution in the form of regular series
\begin{eqnarray}
a(\vx,t,\tau)=\sum_{k=0}^\infty\varepsilon^k a_k(\vx,t,\tau),\quad a_k=\overline{a}_k(\vx,s)+\widetilde{a}_k(\vx,s,\tau),\quad k=0,1,2,\dots
\label{basic-4aa}
\end{eqnarray}
Analytical calculations for each distinguished limit are given in I, they contain the following steps:
(i) writing  the equations for successive approximations, and splitting each equation into its `bar' and `tilde' parts;\ (ii) obtaining the closed systems of equations for the `bar' parts;\ (iii) performing of a large number of integrations by parts and algebraic transformations.  The results of (i)-(iii) give a full solution $a=\overline{a}_k+\widetilde{a}_k$ in any approximation. The above steps can be performed only for distinguished limits; all other asymptotic paths \eqref{paths} produce controversial systems of equations.

\textbf{\emph{The case DL(1)}} with $\overline{\vu}=O(1)\neq 0$  naturally corresponds to the advection speed of order one, hence $s=t$ and the averaged equation of zero approximation is
\begin{eqnarray}\nonumber
\left(\partial_s+ \overline{\vu}\cdot\nabla\right)\overline{a}_{0}=0
\end{eqnarray}
We do not consider this case in detail, all the coefficients in the equations of higher approximations are similar to that in DL(2), one can find it in I.

\textbf{\emph{DL(2) is the most instructive}} and interesting from physical point of view.
Here we have $\overline{\vu}\equiv 0$, hence the speed of
advection is lower than in DL(1) that is described by a longer slow time-scale $s=\varepsilon t$.
The averaged equations of the first three successive approximations are (see Appendix and I, II)
\begin{eqnarray}
&&\left(\partial_s+ \overline{\vV}_0\cdot\nabla\right)\overline{a}_{0}=0,\quad \partial_s\equiv \partial/\partial s\label{4.15}\\
&&\left(\partial_s+ \overline{\vV}_0\cdot\nabla\right)
\overline{a}_{1}+(\overline{\vV}_1\cdot\nabla)\overline{a}_0=0\label{4.16}\\
&&\left(\partial_s+ \overline{\vV}_0\cdot\nabla\right)\overline{a}_{2}+
(\overline{\vV}_1\cdot\nabla)\overline{a}_1+(\overline{\vV}_2\cdot\nabla)\overline{a}_0=
\frac{\partial}{\partial x_i}\left(
\overline{\chi}_{ik}\frac{\partial\overline{a}_0}{\partial x_k}\right),
\label{4.17}
\end{eqnarray}
with notations
\begin{eqnarray}
&&\overline{\vV}_2\equiv\frac{1}{4}\langle[[\vV_0,\widetilde{\vxi}],\widetilde{\vxi}]\rangle +
\frac{1}{2}\langle[\widetilde{\vV}_0,\widetilde{\vV}_0^\tau]\rangle+
\frac{1}{2}\langle[\widetilde{\vxi},\widetilde{\vxi}_s]\rangle
+\frac{1}{2}\langle\widetilde{\vxi}\Div\widetilde{\vu}'+ \widetilde{\vu}'\Div\widetilde{\vxi}\rangle,\label{4.19}
\\
&&\widetilde{\vu}'\equiv\widetilde{\vxi}_s-[\overline{\vV}_0,\widetilde{\vxi}],
\label{4.20a}\\
&&2\overline{\chi}_{ik}\equiv\langle\widetilde{u'}_i\widetilde{\xi}_k+\widetilde{u'}_k\widetilde{\xi}_i\rangle=
\mathfrak{L}_{\overline{\vV}_0}\langle\widetilde{\xi}_i\widetilde{\xi}_k\rangle,\label{4.20}\\
&&\mathfrak{L}_{\overline{\vV}_0}\overline{f}_{ik}\equiv
\left(\partial_s+
\overline{\vV}_0\cdot\nabla\right)\overline{f}_{ik}-\frac{\partial\overline{V}_{0k}}{\partial x_m}\overline{f}_{im}-
\frac{\partial\overline{V}_{0i}}{\partial x_m}\overline{f}_{km}
\label{4.20b}
\end{eqnarray}
where the operator $\mathfrak{L}_{\overline{\vV}_0}$ is such that $\mathfrak{L}_{\overline{\vV}_0}
\overline{f}_{ik}=0$ represents the condition for tensorial field $\overline{f}_{ik}(\vx,t)$
to be `frozen' into $\overline{\vV}_0(\vx,t)$ ($\mathfrak{L}_{\overline{\vV}_0}$ is also known as a Lie derivative).
The summation convention is in use in this paper.

Three equations (\ref{4.15})-(\ref{4.17}) can be written
as a single advection-`diffusion' equation (with an error ${O}(\varepsilon^3)$)
\begin{eqnarray}
&&\left(\partial_s+ \overline{\vV}\cdot\nabla\right)\overline{a} =
\frac{\partial}{\partial x_i}\left(\overline{\kappa}_{ik}\frac{\partial\overline{a}}{\partial x_k}\right)
\label{4.21}\\
&&\overline{\vV}=\overline{\vV}_0+\varepsilon\overline{\vV}_1+\varepsilon^2
\overline{\vV}_2,\label{4.22}\\
&&\overline{\kappa}_{ik}=\varepsilon^2\overline{\chi}_{ik}\label{4.23}\\
&&\overline{a}=\overline{a}_0+\varepsilon\overline{a}_1+\varepsilon^2\overline{a}_2
\label{4.22aa}
\end{eqnarray}
Eqn. (\ref{4.21}) shows that the averaged motion represents a drift with velocity $\overline{\vV}+{O}(\varepsilon^3)$ and
\emph{vibrodiffusion } with matrix coefficients
$\overline{\kappa}_{ik}+{O}(\varepsilon^3)$.
We have introduced the term  {\emph{vibrodiffusion }} on the grounds of the following arguments:
(i) the evolution of $\overline{a}$ is described by an advection-`diffusion'-type equation (\ref{4.21}),
where the matrix of `diffusion' coefficients can be positive or negative (corresponding to diffusion or `antidiffusion');\
(ii) the `diffusion' type term represents a
known source-type term in the second approximation;\
(iii) the equation (\ref{4.21}) is valid only for the regular
asymptotic expansions (\ref{4.22})-(\ref{4.22aa});\ (iv) in I, II we have used the term \emph{pseudoduffusion}, which here is replaced with \emph{vibrodiffusion}
since the former term has already been intensely   used (see Google) for various completely different phenomena.

\textbf{\emph{For DL(3)}} we impose a restriction $\overline{\vV}_0\equiv 0$ and derive equations (see Appendix and I, II):
\begin{eqnarray}
&&\left(\partial_s+ \overline{\vV}_1\cdot\nabla\right)\overline{a}_{0}=0\label{4.15a}\\
&&\left(\partial_s+ \overline{\vV}_1\cdot\nabla\right)\overline{a}_{1}+
(\overline{\vV}_2\cdot\nabla)\overline{a}_0=
\frac{\partial}{\partial x_i}\left(
\overline{\chi}_{ik}\frac{\partial\overline{a}_0}{\partial x_k}\right),
\label{4.17a}\\
&&\overline{\vV}_2\equiv\frac{1}{4}\langle[[\widetilde{{\vV}}_0,\widetilde{\vxi}],\widetilde{\vxi}]\rangle +
\frac{1}{2}\langle[\widetilde{\vV}_0,\widetilde{\vV}_0^\tau]\rangle+
\frac{1}{2}\langle[\widetilde{\vxi},\widetilde{\vxi}_s]\rangle
+\frac{1}{2}\partial_s\langle\widetilde{\vxi}\Div\widetilde{\vxi}\rangle,\label{4.19a}\\
&&2\overline{\chi}_{ik}=
\partial_s\langle\widetilde{\xi}_i\widetilde{\xi}_k\rangle\label{4.20bb}
\end{eqnarray}
\textbf{\emph{For DL(4)}} we impose two restrictions  $\overline{\vV}_0\equiv 0$ and $\overline{\vV}_1\equiv 0$ and derive the equation (see Appendix and I, II)
\begin{eqnarray}
&&\left(\partial_s+ \overline{\vV}_2\cdot\nabla\right)\overline{a}_{0}=
\frac{\partial}{\partial x_i}\left(
\overline{\chi}_{ik}\frac{\partial\overline{a}_0}{\partial x_k}\right),
\label{4.17c}
\end{eqnarray}
with the same $\overline{\vV}_2$ and $\overline{\chi}_{ik}$ as in DL(3).
The comparison between the averaged equations for DL(1)--DL(4) shows that the same coefficients in DL($n$) appear in higher approximations in DL($n-1$)  in the equations of the next order in $\varepsilon$.
The higher approximations DL(5), \emph{etc.}) can be derived similarly, however the calculations become extremely cumbersome.

One can see, that in all presented above distinguished limit solutions the meaning of key \emph{parameter $\varepsilon$ is not uniquely  defined in terms of $\omega$}.
In  DL(2)  $\varepsilon_1=\varepsilon$, $\varepsilon_2=\varepsilon^2$  in (\ref{exact-6a}),
which links $\alpha$, $\beta$ and $\varepsilon$ as:
\begin{equation}
\alpha=1-2\beta,\quad \beta=(1-\alpha)/2,\quad \varepsilon=1/\omega^{(\alpha+1)/2}.
\label{main-eq111}
\end{equation}
Hence, any distinguished limit solution of (\ref{exact-6a}), obtained in terms of $\varepsilon$, produces an infinite number of \emph{parametric solutions}, one solution for any real number $\beta<1$ (or $\alpha>-1$).
Those parametric solutions are mathematically similar but physically different, since they correspond to different magnitudes of given velocities $\vv=\omega^\beta\vu$ and different slow time variables $s=t/\omega^\alpha$. There are two ways of prescribing $\alpha$ and $\beta$. When the slow time-scale $s$ in $\vv(\vx, s,\tau)$ is given, then it defines $\alpha$ and we have to calculate $\beta$ from \eqref{main-eq111}. Alternatively, when the amplitude of $\vv=\omega^\beta\vu$ is given, then it defines $\beta$ and we have to calculate $\alpha$ from \eqref{main-eq111}. Some interesting sets of $\alpha$ and $\beta$ are:

(i) If $\widetilde{\vu}$ is given as the function of variables $\tau=\omega t$ and $s=t$, then  $\alpha=0$, $\beta=1/2$; hence a dimensionless velocity $\vv\sim O(\sqrt{\omega})$ (\ref{variables1}), and in \eqref{basic-4aa} $\varepsilon=1/\sqrt\omega$.

(ii) The most frequently considered case is $\vv\sim O(1)$, then $\alpha=1$, $\beta=0$, $s=t/\omega$, and $\varepsilon=1/\omega$.

 (iii) Rather exotic possibility corresponds to the case $\alpha=\beta =1/3$. Here, $s=t/\sqrt[3]\varepsilon$, $\varepsilon=\omega^{-2/3}$ and velocity $\vv=O(\sqrt[3]\omega)$. Such a scaling is required if a particular slow time-scale $s=t/\sqrt[3]\omega$ is prescribed in $\widetilde{\vu}$.

 (iv) The above results for DL(2) are particularly striking for the case of  $\vu=\widetilde{\vu}(\vx,\tau)$, which is independent of $s$ and represents a flow with purely periodic velocity oscillations. An intricate question arises: how to choose any particular scale of $s$?
 Our answer is: in this case the slow time-scale is uniquely determined by the magnitude $O(\omega^\beta)$ of $\vv$. For every value of $\beta$ we obtain the related slow time $s=t/\omega^{(1-2\beta)}$. It should be accepted that solutions with different $\beta$ are physically different since they correspond to different orders of prescribed velocity field $\vv$.

 (v) The functional dependence \eqref{main-eq111} between $\alpha$ and $\beta$ is physically natural: the increasing of the amplitude of a given velocity (increasing $\beta$) leads to the decreasing of the slow time-scale (decreasing $\alpha$).
In general, transformations similar to (\ref{main-eq111}) produce an infinite number of \emph{parametric solutions} for each case DL(1-4).

\section{Examples}

The above results are obtained for arbitrary function $\widetilde{\vu}(\vx, s,\tau)$. Hence the number of interesting examples is infinite. Let us consider five instructive classes of $\widetilde{\vu}(\vx, s,\tau)$.

\textbf{\emph{Example 1.} \emph{The superposition of two modulated oscillatory fields}}
\begin{eqnarray}
&&\widetilde{\vu}(\vx,s,\tau)=\overline{\vp}(\vx,s)\sin\tau+
\overline{\vq}(\vx,s)\cos\tau  \label{Example-1-1}
\end{eqnarray}
where
$\overline{\vp}$ and $\overline{\vq}$ are arbitrary bar-functions. The straightforward calculations yield
$
[\widetilde{\vu},\widetilde{\vxi}]=[\overline{\vp},\overline{\vq}].
$
Equations (\ref{4.18}), (\ref{4.19}), \eqref{4.20} yield
\begin{eqnarray}
&&\overline{\vV}_0=\frac{1}{2}[\overline{\vp},\overline{\vq}],\quad
\overline{\vV}_1\equiv 0,
\label{Example-1-4}\\
&&\overline{\vV}_2=\frac{1}{8}\left([\overline{\vP},\overline{\vp}]+[\overline{\vQ},\overline{\vq}]\right)
-\frac{1}{4}\left([\overline{\vp}_s,\overline{\vp}]+[\overline{\vq}_s,\overline{\vq}]\right)+\label{Example-1-4a}\\
&&+\frac{1}{4}\left(\overline{\vp}\,\Div\overline{\vP}'+\overline{\vq}\Div\overline{\vQ}'
+\overline{\vP}'\Div\overline{\vp}+\overline{\vQ}'\Div\overline{\vq}\right),\nonumber\\
&&\overline{\vP}\equiv[\overline{\vV}_0,\overline{\vp}],\quad
\overline{\vQ}\equiv[\overline{\vV}_0,\overline{\vq}],\quad
\overline{\vP}'\equiv\overline{\vp}_s-\overline{\vP},\quad
\overline{\vQ}'\equiv\overline{\vq}_s-\overline{\vQ},\nonumber\\
&&\langle\widetilde{\xi}_i\widetilde{\xi}_k\rangle=
\frac{1}{2}(\overline{p}_i \overline{p}_k + \overline{q}_i \overline{q}_k)\label{Example-1-5}
\end{eqnarray}
Expression for the \emph{vibrodiffusion} matrix $\overline{\kappa}_{ik}$ follows after the substitution of (\ref{Example-1-5}) into
(\ref{4.20}). The expression $\widetilde{\vu}$ (\ref{Example-1-1}) is general enough to produce any given function
$\overline{\vV}_0(\vx,t)$. Indeed, in order to obtain $\overline{\vp}(\vx,t)$ and $\overline{\vq}(\vx,t)$ one has to solve the equation
\begin{eqnarray}\label{bi-linear}
[\overline{\vp},\overline{\vq}]=(\overline{\vq}\cdot\nabla)\overline{\vp}-
(\overline{\vp}\cdot\nabla)\overline{\vq}=2\overline{\vV}_0(\vx,s)
\end{eqnarray}
which represents an undetermined bi-linear PDE-problem for two unknown functions.

\textbf{\emph{Example 2.} \emph{ Stokes drift}}.

The dimensionless plane velocity field is (see \cite{Stokes, Lamb, Debnath})
\begin{eqnarray}
\widetilde{\vu}=Ae^{ky}\left(\begin{array} {c} \cos(kx-\tau)\\ \sin (kx-\tau)\end{array}\right),\quad (x,y)\equiv(x_1,x_2)
\label{Example-3-1}
\end{eqnarray}
where one can choose $A=1$ and $k=1$; however, we keep both $A$ and $k$ in the formulae in order to trace the
physical meaning. The fields $\overline{\vp}(x,y)$, $\overline{\vq}(x,y)$ (\ref{Example-1-1}) are
\begin{eqnarray}
\overline{\vp}=Ae^{ky}\left(\begin{array}{c} \sin kx\\ -\cos kx\end{array}\right),\quad
\overline{\vq}=Ae^{ky}\left(\begin{array}{c} \cos kx\\ \sin kx\end{array}\right)
\label{Example-3-2}
\end{eqnarray}
The calculations of (\ref{Example-1-4}) yield
\begin{eqnarray}
\overline{{\vV}}_0=k A^2 e^{2ky}\left(\begin{array}{c} 1\\ 0\end{array}\right), \quad  \overline{{\vV}}_1\equiv 0
\label{Example-3-3}
\end{eqnarray}
which represents the classical Stokes drift and the first correction
to it (which vanishes).  For brevity, the explicit formula for $\overline{{\vV}}_2$  is not given here.
Further calculations show that
\begin{eqnarray}\nonumber
&&\overline{\chi}_{ik}=-\overline{\chi}
\begin{pmatrix}
0 & 1\\
1 & 0
\end{pmatrix},\quad\text{with}\quad \overline{\chi}\equiv\frac{1}{4}k^2 A^4 e^{3ky}
\end{eqnarray}
One can see that the eigenvalues $\overline{\chi}_1=-\overline{\chi}$ and $\overline{\chi}_2=\overline{\chi}$
correspond to strongly anisotropic \emph{vibrodiffusion}.
The averaged equation (\ref{4.21}) (with an error $O(\varepsilon^3)$) can be written as
\begin{eqnarray}
&&\overline{a}_{t}+(\overline{V}_0+\varepsilon^2 \overline{V}_2) \overline{a}_x=
    \varepsilon^2(\overline{\chi}_y \overline{a}_{x}+\overline{\chi}\, \overline{a}_{xy})\label{Example-3-5}\\
    &&\overline{a}=\overline{a}_0+\varepsilon\overline{a}_1+\varepsilon^2\overline{a}_2\nonumber
\end{eqnarray}
where $\overline{V}_0$ and $\overline{V}_2$ are the $x$-components of corresponding velocities (their $y$-components
vanish). This equation has an exact solution $\overline{a}=\overline{a}(y)$ where $\overline{a}(y)$ is an arbitrary function, which is not affected  by \emph{vibrodiffusion}.

\textbf{\emph{Example 3.}  \emph{A spherical `acoustic' wave.}}

A velocity potential for an outgoing spherical wave is
\begin{eqnarray}
&& \widetilde{\phi}=\frac{A}{r}\sin(kr-\tau)\label{Example-4-1}
\end{eqnarray}
where $A$, $k$, and $r$ are amplitude, wavenumber, and radius in a spherical coordinate system. The velocity is
purely radial and has a form (\ref{Example-1-1})
\begin{eqnarray}
&& \widetilde{u}=\overline{p}\sin\tau+\overline{q}\cos\tau,\\
&&\overline{p}=A\left(\frac{1}{r^2}\cos kr+\frac{k}{r}\sin kr\right),\quad\overline{q}=A\left(-\frac{1}{r^2}\sin
kr+\frac{k}{r}\cos kr\right)\label{Example-4-1a}
\end{eqnarray}
where $\widetilde{u}, \overline{p}$, and $\overline{q}$ are radial components of corresponding vector-fields. The
fields $\widetilde{\vxi}$ and $[\widetilde{\vu},\widetilde{\vxi}]$ are also purely radial; the radial component for the
commutator is
\begin{eqnarray}
&& \widetilde{\xi} \widetilde{u}_r-\widetilde{u}\widetilde{\xi}_r=A^2 k^3/r^2\label{Example-4-3}
\end{eqnarray}
where $\widetilde{\xi}$ is the radial component of $\widetilde{\vxi}$ and subscript $r$ stands for the radial derivative. The drift
(\ref{Example-1-4}) is purely radial with
\begin{eqnarray}
&&\overline{V}_0=\frac{A^2 k^3}{2\,r^2},\quad
\overline{V}_1=0,\quad\overline{V}_2=\frac{A^4k^5}{16r^4}\left(3k^2-\frac{5}{r^2}\right)
\label{Example-4-5}
\end{eqnarray}
It is interesting that $\overline{V}_0$ formally coincides with the velocity, caused by a point source in an
incompressible fluid, and (for small $r$) the value of  $\overline{V}_2$
dominates over $\overline{V}_0$, so the series is likely to be diverging. Further calculations yield
\begin{eqnarray}
&&\langle\xi^2\rangle=\frac{A^2}{2r^2}(k^2+1/r^2),\quad \overline{\chi}=A^4k^5/4r^2>0
\label{Example-4-6}
\end{eqnarray}
where $\overline{\chi}$ stands for the only nonzero $rr$-component of $\overline{\chi}_{ik}$. One can see that in this
case \emph{vibrodiffusion} appears as ordinary diffusion.

\textbf{\emph{Example 4.}  \emph{ $\overline{\vV}_1$-drift.}}

It is interesting to consider such flows for which $\overline{\vV}_0\equiv 0$ but $\overline{\vV}_1\neq 0$. Let a velocity field
be a superposition of two standing waves of frequencies $\omega$ and $2\omega$:
\begin{eqnarray}
&&\widetilde{\vu}(\vx,t,\tau)=\overline{\vp}(\vx,t)\sin\tau+\overline{\vq}(\vx,t)\sin 2\tau  \label{Example-5-1}\\
&&[\widetilde{\vu},\widetilde{\vxi}]=\frac{1}{2}[\overline{\vp},\overline{\vq}](2\cos\tau\sin 2\tau-\cos
2\tau\sin\tau)\label{Example-5-3}
\end{eqnarray}
Hence (\ref{4.18}) yields
\begin{eqnarray}\label{Example-5-4}
\overline{\vV}_0=\frac{1}{2}\langle[\widetilde{\vu},\widetilde{\vxi}]\rangle\equiv 0,\quad
\overline{\vV}_1=\frac{1}{3}\langle[[\widetilde{\vu},\widetilde{\vxi}],\widetilde{\vxi}]\rangle=
\frac{1}{8}[[\overline{\vp},\overline{\vq}],\overline{\vp}]
\end{eqnarray}
These expressions produce infinitely many examples of the flows with $\overline{\vV}_1$-drift. At the same time it shows that for a standing wave the $\overline{\vV}_1$-drift is absent.

\textbf{\emph{Example 5}  \emph{ Chaotic dynamics for $\overline{\vV}_0$.}}

Let a solenoidal/incompressible velocity (\ref{Example-1-1})
be
\begin{eqnarray}
\overline{\vp}=\left(\begin{array}{c}
  \cos y \\
  0 \\
  \sin y
\end{array}\right), \qquad
\overline{\vq}=\left(\begin{array}{c}
  a\sin z \\
  b\sin x+a\cos z \\
  b\cos x
\end{array}\right)
\end{eqnarray}
where  $(x,y,z)$ are  cartesian coordinates; $a, b$ are constants. Either of these fields, taken separately, produces
simple integrable dynamics of particles. The calculations yield
\begin{eqnarray}
\overline{\vV}_0=\left(\begin{array}{c}
  -a\sin y\sin x-2b\sin y\cos z \\
  b\sin z\sin y-a\cos x\cos y \\
  b\cos z\cos y+2a\sin x\cos y
\end{array}\right),
\end{eqnarray}
The computations of lagrangian dynamics for $d\overline{\vx}_0/ds=\overline{\vV}_0$ (based on (\ref{exact-1a}) for DL(2), see I) were performed by Prof. A.B.Morgulis (private communications). He has shown that this steady averaged flow exhibits chaotic dynamics of particles.
In particular, positive
Lyapunov exponents have been observed. Hence, the drift created by a simple oscillatory field can
produce complex averaged lagrangian dynamics.

\section{Discussion}

1. Our study is aimed to create a general viewpoint on distinguished limits, drifts, and vibrodiffusion based on the rigorous implementation of the two-timing method. We do not use any additional suggestions and assumptions, hence any failure to interpret our results (either physically or mathematically) can be explained only by insufficiency of two time-scales. Indeed, the presence of scaling parameters, such as $St$, $\varepsilon_1$, and $\varepsilon_2$, allows to introduce an infinite number of additional time-scales.
From this perspective the problem we study can be seen as a test for sufficiency of the two-timing method.
In particular, some secular (in $s$) terms could appear in solutions of \eqref{4.17} and \eqref{4.17a}.
If it is proven and recognised as unacceptable, then one can suggest that the two-timing method fails at the orders of approximations, where  \emph{vibrodiffusion} appears.
In this case further time-scales must be introduced, which require the systematic development of three-timing \emph{etc.} methods. However, any such systematic development which allows to derive the averaged equations with three or more time scales from an original PDE is still unknown.

2. The two-timing method has been used by many authors, see \cite{Kevorkian, Nayfeh, Verhulst}, however our analysis goes well beyond the usual calculations of the main approximation in various special cases.
Our analytic calculations are straightforward by their nature, but they do include a large number of integration by parts and algebraic transformations;
the performing of such calculations in a general formulation represents a `champion-type' result by itself.
These calculations are described in detail in the Appendix and in the arXiv papers I, II.

3. A different (from the presented in this paper approach) is given in \cite{Yudovich, VladimirovDr1} where \emph{an inspection procedure} for calculation of distinguished paths is proposed and the concept of `strong',  `moderate', and `weak' oscillations (or vibrations) is introduced and exploited.
The advantage of our approach is the additional possibility to vary the slow time-scale $s$. It makes the structure of asymptotic solutions more flexible, and allows us to consider the settings of small parameters which are more `natural' physically.
Say, in the physically `natural' case $\vv\sim O(1)$ in \eqref{variables1}   we should take $\alpha=1$, $\beta=0$, $s=t/\omega$, and $\varepsilon=1/\omega$ \eqref{main-eq111}. However, if we take $s=t$ (which also could be seen as physically `natural'), then we are forced to chose $\vv\sim O(\sqrt{\omega})$ that could be viewed as physically `artificial'. Such an interplay of scales may create confusion and misunderstanding, but it is certainly useful for considering applications.

4. \emph{Vibrodiffusion} in  DL(1-4) is especially interesting to study.
Its physical mechanism is discussed in I, II.
It is worth to emphasize that a \emph{vibrodiffusion} matrix/coefficient for different flows can be positive, negative, or can change its sign in space and time.
At the same time the appearance of vibrodiffusion as a \emph{Lie-derivative of an averaged quadratic displacement tensor} \eqref{4.20}, \eqref{4.20b} still requires physical explanation.
The most important open question is the possibility of the secular growth (in $s$) of averaged solutions due to vibrodiffusion. Such secular terms can appear for the flows with the vibrodiffusion matrix $\overline{\chi}_{ik}$ \eqref{4.20} monotonically increasing in $s$ and are unlikely to appear for the oscillating $\overline{\chi}_{ik}$ in $s$ matrix. If such a growth does appear for constant or oscillating $\overline{\chi}_{ik}$, then two time-scales are not sufficient one has to introduce three or more time-scales.

5. The results for DL(2) \eqref{main-eq111}  are particularly striking for the case of  $\vu=\widetilde{\vu}(\vx,\tau)$, which is independent of $s$ and represents a flow with purely periodic velocity oscillations. Here, we give the answer to an intricate question: how to choose any particular scale of $s$?
Our answer is: the slow time-scale is uniquely determined by the magnitude $O(\omega^\beta)$ of $\vv$. For every value of $\beta$ we obtain the related slow time $s=t/\omega^{(1-2\beta)}$. It should be accepted that solutions with different $\beta$ are physically different since they correspond to different orders of the prescribed velocity field $\vv$.

6. One can rewrite the approximate solution \eqref{4.22aa} (along with its tilde-parts given in I, II) back to variable $t$ and substitute it into the exact original equation (\ref{exact-1}); then a small residual (a nonzero right-hand-side in (\ref{exact-1})) appears.
The method used allows to produce an approximate solution with a residual
as small as needed. However, the next logical step is more challenging: one has to prove that a solution with a
small residual is close to the exact one.
Such proofs had been performed by \cite{Simonenko, Levenshtam} for the case of vibrational convection. Similar justifications
for other oscillatory flows are not available yet.

7. All results of this paper have been obtained for the class of $\tau$-periodic functions, which is self-consistent. One can consider more general
classes of quasi-periodic, non-periodic, or chaotic solutions. The discussion on this topic is given in I, II.
However it is worth to understand the properties of periodic oscillations, in order not to link these properties exclusively to more general solutions.
On the other hand, the relative simplicity of calculations for the  $\tau$-periodic solutions allows to obtain more advanced results (than, say, for a chaotic solution). In this case our results can serve as a heuristic guidance for making assumptions on the properties of more general solutions.

8. \emph{Example 5} demonstrates that a drift can produce chaotic averaged dynamics of particles.
This result leads to numerous new questions and opportunities such as: (i) what is
the relationship between chaotic motions for an original dynamical system and for the averaged one?  (ii) how can a \emph{chaotic drift} and  \emph{vibrodiffusion} complement each other?
(iii) how a chaotic drift can be used in the theory of mixing? (iv) since the averaged dynamics is chaotic, then how the related results by
\cite{Arnold,Aref,Ottino,Wiggins,Chierchia} and many others can be used efficiently.

9. All \emph{Examples} in Sect. 4 are presented very briefly. For a systematic study, each example requires a separate paper. More examples are given in I and II.

10. It is worth to calculate the characteristics/tragectories directly by solving (\ref{exact-1a}) with the use of the same two-timing method and to compare the related solutions
for drifts with the results presented above. Such calculations are presented in I, II.

11. Similar results for a passive vectorial admixture are also presented in I, II.
They are closely linked to the problem of \emph{kinematic $MHD$-dynamo}, see \cite{Moffatt}.
It is physically apparent, that for the majority of shear drift velocities $\overline{\vV}(\vx,t)$ the averaged stretching of material
lines  produces the linear in $s$ growth of a magnetic field $|\overline{\vh}|\sim t$. At the same time, for the averaged flows with
exponential stretching of averaged material lines these examples will inevitably show the exponential growth.
Another closely related research topic is the `active' advection of a vectorial admixture field (vorticity) in the vortex dynamics of oscillating flows; here the major phenomenon is the striking  \emph{Langmuir circulations}, see
\cite{Craik0}, which has been recently analyzed from a new perspective by \cite{VladimirovL, VladProc}. Both those topics are worth to be studied by the above approach.

\begin{acknowledgments}
The author is
grateful to Profs. K.I.Ilin, M.E.McIntyre,  A.B.Morgulis, T.J.Pedley, M.R.E.Proctor, and D.W.Hughes  for helpful discussions.
Special thanks to Profs. A.D.D.Craik and H.K.Moffatt for reading this manuscript and making important and useful critical remarks.
\end{acknowledgments}

\appendix

\section{DL(2): Asymptotic Procedure and Solution.  \label{sect04}}

First, we list some properties of $\tau$-differentiation and tilde-integration \eqref{ti-integr} which are intensely used in calculations below.
For $\tau$-derivatives it is clear that
\begin{eqnarray}
f_\tau=\overline{{f}}_\tau+\widetilde{f}_\tau=\widetilde{f}_\tau, \quad
\langle f_\tau\rangle=\langle\widetilde{f}_\tau\rangle=0 \label{oper-6}
\end{eqnarray}
The product of two tilde-functions $\widetilde{f}$ and $\widetilde{g}$ forms a general function:
$\widetilde{f}\widetilde{g}\equiv F$, say. Separating tilde-part $\widetilde{F}$ from $F$ we
write
\begin{eqnarray}
&&\widetilde{F}=F-\langle F\rangle
=\widetilde{f}\widetilde{g}-\langle\widetilde{f}\widetilde{g}\rangle=
\widetilde{\widetilde{f}\widetilde{g}}\equiv\{\widetilde{f}\widetilde{g}\}
\label{oper-5}
\end{eqnarray}
Since we do not use a two-level tilde notation for the tilde-parts of long expressions, we introduce braces
instead. As the average operation is proportional to the integration over $\tau$, for products
containing tilde-functions $\widetilde{f},\widetilde{g},\widetilde{h}$ and their derivatives we have
\begin{eqnarray}
&&\langle\widetilde{f}\widetilde{g}_\tau\rangle=\langle(\widetilde{f}\widetilde{g})_\tau\rangle-
\langle\widetilde{f}_\tau\widetilde{g}\rangle=-\langle\widetilde{f}_\tau \widetilde{g}\rangle=-
\langle\widetilde{f}_\tau g\rangle
\label{oper-9}\\
&&\langle\widetilde{f}_\tau\widetilde{g}\widetilde{h}\rangle+\langle\widetilde{f}\widetilde{g}_\tau\widetilde{h}\rangle+
\langle\widetilde{f}\widetilde{g}\widetilde{h}_\tau\rangle=0
\label{oper-9a}\\
&&\langle\widetilde{f}\widetilde{g}^\tau\rangle=\langle(\widetilde{f}^\tau\widetilde{g}^\tau)_\tau\rangle-
\langle\widetilde{f}^\tau\widetilde{g}\rangle=-\langle\widetilde{f}^\tau \widetilde{g}\rangle=-
\langle\widetilde{f}^\tau g\rangle
\label{oper-10}
\end{eqnarray}
which represent different versions of the integration by parts. Similarly, for commutators we have
\begin{eqnarray}
&&\langle[\widetilde{\va},\widetilde{\vb}_\tau]\rangle=-\langle[\widetilde{\va}_\tau,\widetilde{\vb}]\rangle=-
\langle[\widetilde{\va}_\tau, \vb]\rangle,\
\langle[\widetilde{\va},\widetilde{\vb}^\tau]\rangle=-\langle[\widetilde{\va}^\tau,\widetilde{\vb}]\rangle=-
\langle[\widetilde{\va}^\tau, \vb]\rangle
\label{oper-15}
\end{eqnarray}
Now we can describe the obtaining of solution to equation \eqref{exact-6c} for $n=2$
\begin{eqnarray}
&& a_\tau+\varepsilon^2 a_s+\varepsilon(\widetilde{\vu}\cdot\nabla)\,a=0
\label{Abasic-2}
\end{eqnarray}
The substitution of (\ref{basic-4aa}) into this equation produces the equations of successive approximations
\begin{eqnarray}
&&a_{0\tau}=0 \label{Abasic-5}\\
&&a_{1\tau}=-(\widetilde{\vu}\cdot\nabla)\, a_0\label{Abasic-6}\\
&&a_{n\tau}=-(\widetilde{\vu}\cdot\nabla)\,
a_{n-1}-\partial_s\, a_{n-2},\quad\partial_s\equiv\partial/\partial s,\quad n=2,3,4,\dots
\label{Abasic-7}
\end{eqnarray}
After calculations, the tilde-parts $\widetilde{a}_k$, $k=0,1,2,3,4$ are given by the explicit recurrent expressions
\begin{eqnarray}
&&\widetilde{a}_0\equiv 0,\label{4.11a}\\
&& \widetilde{a}_1= -(\widetilde{\vxi}\cdot\nabla)\,\overline{a}_0,\label{A4.11}\\
&&\widetilde{a}_{2}=-(\widetilde{\vxi}\cdot\nabla)\, \overline{a}_1
-\{(\widetilde{\vu}\cdot\nabla)\,\widetilde{a}_1\}^\tau, \label{A4.12}\\
&&\widetilde{a}_{3}=-(\widetilde{\vxi}\cdot\nabla)\,\overline{a}_2
-\{(\widetilde{\vu}\cdot\nabla)\,\widetilde{a}_2\}^\tau-\widetilde{a}_{1s}^\tau, \label{A4.13}\\
&&\widetilde{a}_{4}=-(\widetilde{\vxi}\cdot\nabla)\,\overline{a}_3-
\{(\widetilde{\vu}\cdot\nabla)\,\widetilde{a}_3\}^\tau-\widetilde{a}_{2s}^\tau, \label{A4.14}
\end{eqnarray}
Further calculations show that the bar-parts $\overline{a}_k$ satisfy the equations \eqref{4.15}-\eqref{4.20b}.
Let us present the derivations of \eqref{4.11a}-\eqref{A4.14} and \eqref{4.15}-\eqref{4.20b}.

\underline{\emph{The zero-order equation}}  (\ref{Abasic-5})  is:
\begin{eqnarray}
&&a_{0\tau}=0\label{A01-appr-1}
\end{eqnarray}
The substitution of $a_{0}=\overline{a}_0(\vx,s)+\widetilde{a}_0(\vx,s,\tau)$ into (\ref{A01-appr-1}) gives
$\widetilde{a}_{0\tau}= 0$. Its tilde-integration \eqref{ti-integr}   produces a unique (inside the
tilde-class) solution $\widetilde{a}_0\equiv 0$. At the same time (\ref{A01-appr-1}) does not impose any
restrictions on $\overline{a}_0(\vx,t)$, which must be determined from the next approximations. Thus the results
derivable from (\ref{A01-appr-1}) are:
\begin{eqnarray} &&\widetilde{a}_0(\vx,s,\tau)\equiv 0;\quad \overline{a}_0(\vx,s) \ \text{is not defined}\label{A01-appr-1a}
\end{eqnarray}

\underline{\emph{The first-order equation} } (\ref{Abasic-6})  is
\begin{eqnarray}
&&a_{1\tau}=-(\widetilde{\vu}\cdot\nabla)\, a_0 \label{A01-appr-3}
\end{eqnarray}
The use of $\widetilde{a}_0\equiv 0$ and $\overline{a}_{1\tau}\equiv 0$ reduces (\ref{A01-appr-3}) to
the equation $\widetilde{a}_{1\tau}=-(\widetilde{\vu}\cdot\nabla)\,\overline{a}_0$. Its tilde-integration gives the unique solution
\begin{eqnarray}
&&\widetilde{a}_{1}=-(\widetilde{\vxi}\cdot\nabla)\,\overline{a}_0\label{01-appr-7}
\end{eqnarray}
 Hence,
\begin{eqnarray}
&&a_1=\overline{a}_1 -(\widetilde{\vxi}\cdot\nabla)\,\overline{a}_0
\label{01-appr-8}
\end{eqnarray}
where $\overline{a}_0(\vx,t), \overline{a}_1(\vx,t)$ are not defined.

\underline{\emph{The second-order equation}} ((\ref{Abasic-7}) for $n=2$) is
\begin{eqnarray}
&&a_{2\tau}=-(\widetilde{\vu}\cdot\nabla)\,a_1-
a_{0s}\label{2-appr-1}
\end{eqnarray}
The use of (\ref{A01-appr-1a}) and $\overline{a}_{2\tau}\equiv 0$ transforms (\ref{2-appr-1}) to
\begin{eqnarray}
&&\widetilde{a}_{2\tau}=-(\widetilde{\vu}\cdot\nabla)\,\overline{a}_1- (\widetilde{\vu}\cdot\nabla)\,
\widetilde{a}_1-
\overline{a}_{0s}\label{2-appr-2}
\end{eqnarray}
Its bar-part is
\begin{eqnarray}
&&\overline{a}_{0s}=-\langle(\widetilde{\vu}\cdot\nabla)\,\widetilde{a}_1\rangle
\label{2-appr-4}
\end{eqnarray}
where we have used $\langle{\widetilde{a}_{2\tau}}\rangle=0$,
$\langle(\widetilde{\vu}\cdot\nabla)\,\overline{a}_1\rangle=0$, and $\langle
\overline{a}_{0s}\rangle= \overline{a}_{0s}$.
The substitution of (\ref{01-appr-7}) into (\ref{2-appr-4}) produces the equation
\begin{eqnarray}\label{2-appr-5}
&&\overline{a}_{0t}=
\langle(\widetilde{\vu}\cdot\nabla)(\widetilde{\vxi}\cdot\nabla)\rangle\,\overline{a}_0
\end{eqnarray}
One may expect that the RHS of (\ref{2-appr-5}) contains both first and second spatial derivatives of $\overline{a}_0$, however \emph{all second
derivatives vanish}. In order to prove it we introduce a commutator
\begin{eqnarray}
&&\vK\equiv[\widetilde{\vxi},\widetilde{\vu}]=
(\widetilde{\vu}\cdot\nabla)\widetilde{\vxi}-(\widetilde{\vxi}\cdot\nabla)\widetilde{\vu},\label{App1-2}\\
&&\vK\cdot\nabla=(\widetilde{\vu}\cdot\nabla)(\widetilde{\vxi}\cdot\nabla)-(\widetilde{\vxi}\cdot\nabla)(\widetilde{\vu}\cdot\nabla)
\label{App1-1}
\end{eqnarray}
The bar-part of (\ref{App1-1}) is
\begin{eqnarray}
&&\langle(\widetilde{\vu}\cdot\nabla)(\widetilde{\vxi}\cdot\nabla)\rangle
=\langle(\widetilde{\vxi}\cdot\nabla)(\widetilde{\vu}\cdot\nabla)\rangle+\overline{\vK}\cdot\nabla\label{App1-3}
\end{eqnarray}
At the same time the integration by parts over $\tau$ gives
\begin{eqnarray}
&&\langle(\widetilde{\vu}\cdot\nabla)(\widetilde{\vxi}\cdot\nabla)\rangle
=-\langle(\widetilde{\vxi}\cdot\nabla)(\widetilde{\vu}\cdot\nabla)\rangle,\quad\widetilde{\vu}\equiv\widetilde{\vxi}_\tau
\label{App1-4}
\end{eqnarray}
Combining (\ref{App1-3}) and (\ref{App1-4}) we obtain
\begin{eqnarray}
&&\langle(\widetilde{\vu}\cdot\nabla)(\widetilde{\vxi}\cdot\nabla)\rangle
=\frac{1}{2}\overline{\vK}\cdot\nabla\label{App1-5}
\end{eqnarray}
which reduces (\ref{2-appr-5}) to the advection equation (\ref{4.15}) with
\begin{eqnarray}
&&\overline{\vV}_0\equiv-\langle(\widetilde{\vu}\cdot\nabla)\,\widetilde{\vxi}\rangle=
-\frac{1}{2}\langle[\widetilde{\vxi},\widetilde{\vu}]\rangle=-\frac{1}{2}\overline{\vK}
\label{2-appr-8}
\end{eqnarray}
which also gives the main term in drift velocity (\ref{4.18}). The tilde-part of (\ref{2-appr-2}) appears after subtracting (\ref{2-appr-4})
from (\ref{2-appr-2}):
\begin{eqnarray}
&&\widetilde{a}_{2\tau}= -(\widetilde{\vu}\cdot\nabla)\, \overline{a}_1 -
\{(\widetilde{\vu}\cdot\nabla)\,\widetilde{a}_1\}.
\label{2-appr-10}
\end{eqnarray}
Its tilde-integration with the use of (\ref{01-appr-7}) gives (\ref{A4.12}):
\begin{eqnarray}
&&\widetilde{a}_{2}=-(\widetilde{\vxi}\cdot\nabla)\, \overline{a}_1 +
\{(\widetilde{\vu}\cdot\nabla)(\widetilde{\vxi}\cdot\nabla)\}^\tau\overline{a}_0
\label{2-appr-11}
\end{eqnarray}
Hence, $a_2$ can be written as
\begin{eqnarray}
&&a_2=\overline{a}_2+\widetilde{a}_2
\quad
\label{2-appr-12}
\end{eqnarray}
where $\overline{a}_0$ and $\widetilde{a}_{2}$ are given by (\ref{4.15}), (\ref{2-appr-11}), while $\overline{a}_1,\overline{a}_2$ are not defined.

\underline{\emph{The third-order equation}} ((\ref{Abasic-7}) for $n=3$) is:
\begin{eqnarray}
&&\widetilde{a}_{3\tau}=-(\widetilde{\vu}\cdot\nabla)\, a_2-a_{1s}\label{3-appr-1}
\end{eqnarray}
Its bar-part is
\begin{eqnarray}
&&\overline{a}_{1s}= -\langle(\widetilde{\vu}\cdot\nabla)\,\widetilde{a}_2\rangle.
\label{3-appr-2}
\end{eqnarray}
The substitution of (\ref{2-appr-11}) into (\ref{3-appr-2}), the use of  $\widetilde{\vu}\equiv\widetilde{\vxi}_\tau$,
and the integration by parts yield
\begin{eqnarray}
&&\overline{a}_{1s}=
\langle(\widetilde{\vu}\cdot\nabla)(\widetilde{\vxi}\cdot\nabla)\rangle\overline{a}_1+
\langle(\widetilde{\vxi}\cdot\nabla)(\widetilde{\vu}\cdot\nabla)(\widetilde{\vxi}\cdot\nabla)\rangle\overline{a}_0
\label{3-appr-3}
\end{eqnarray}
where $\langle(\widetilde{\vu}\cdot\nabla)(\widetilde{\vxi}\cdot\nabla)\rangle$ has been already simplified in
(\ref{App1-5}). The second term in the RHS of (\ref{3-appr-3}) formally contains the third, second, and first spatial
derivatives of $\overline{a}_0$; however \emph{all the third  and  second derivatives vanish}. To prove it, first, we
use (\ref{oper-9a}):
\begin{eqnarray}
&&\langle(\widetilde{\vxi}\cdot\nabla)(\widetilde{\vu}\cdot\nabla)(\widetilde{\vxi}\cdot\nabla)\rangle=
-\langle(\widetilde{\vu}\cdot\nabla)(\widetilde{\vxi}\cdot\nabla)(\widetilde{\vxi}\cdot\nabla)\rangle-
\langle(\widetilde{\vxi}\cdot\nabla)(\widetilde{\vxi}\cdot\nabla)(\widetilde{\vu}\cdot\nabla)\rangle\label{App2-1}
\end{eqnarray}
Then we use (\ref{App1-2}), (\ref{App1-1}) to transform the sequence of operators $(\widetilde{\vxi}\cdot\nabla)$ and
$(\widetilde{\vu}\cdot\nabla)$ in each term in the RHS of (\ref{App2-1}) into their sequence in the LHS. The result is
\begin{eqnarray}
&&\langle(\widetilde{\vxi}\cdot\nabla)(\widetilde{\vu}\cdot\nabla)(\widetilde{\vxi}\cdot\nabla)\rangle=
\frac{1}{3}\overline{\vK'}\cdot\nabla,\quad \vK'\equiv[\vK,\widetilde{\vxi}]\label{App2-2}
\end{eqnarray}
As the result (\ref{3-appr-3}) takes form (\ref{4.16})  with $\overline{\vV}_0$ (\ref{2-appr-8}) and
\begin{eqnarray}
&&\overline{\vV}_1\equiv-\langle(\widetilde{\vxi}\cdot\nabla)(\widetilde{\vu}\cdot\nabla)\widetilde{\vxi})\rangle=
-\frac{1}{3}\langle[[\widetilde{\vxi},\widetilde{\vu}],\widetilde{\vxi}]\rangle=-\frac{1}{3}\overline{\vK'}
\label{3-appr-4a}
\end{eqnarray}
which gives (\ref{4.18}). The tilde-part of (\ref{3-appr-1}) after its integration gives
\begin{eqnarray}
&&\widetilde{a}_{3}=-(\widetilde{\vxi}\cdot\nabla)\,\overline{a}_2-
\{(\widetilde{\vu}\cdot\nabla)\,\widetilde{a}_2\}^\tau-\widetilde{a}_{1t}^\tau,
\quad\widetilde{a}_1^\tau=-(\widetilde{\vxi}^\tau\cdot\nabla)\, \overline{a}_0 \label{3-appr-5}
\end{eqnarray}
where $\widetilde{a}_2$ is given by (\ref{2-appr-11}).
Hence, $a_3$ can be written as
\begin{eqnarray}
&&a_3=\overline{a}_3+\widetilde{a}_3
\label{3-appr-6}
\end{eqnarray}
where $\overline{a}_0$, $\overline{a}_1$, $\widetilde{a}_{2}$, and $\widetilde{a}_{3}$ are given by (\ref{4.15}),
(\ref{4.16}), (\ref{2-appr-11}), and (\ref{3-appr-5}), while $\overline{a}_2,\overline{a}_3$ are not defined.

\underline{\emph{The fourth-order equation}} ((\ref{Abasic-7}) for $n=4$) is:
\begin{eqnarray}
&&\widetilde{a}_{4\tau}=-(\widetilde{\vu}\cdot\nabla)\, a_3-a_{2s}\label{4-appr-1}
\end{eqnarray}
Its bar-part is
\begin{eqnarray}
&&\overline{a}_{2s}= -\langle(\widetilde{\vu}\cdot\nabla)\,\widetilde{a}_3\rangle
\label{4-appr-2}
\end{eqnarray}
The substitution of $\widetilde{a}_3$ (\ref{3-appr-5})  into (\ref{4-appr-2}), $\widetilde{a}_2$ (\ref{2-appr-11}) into
$\widetilde{a}_3$ (\ref{3-appr-5}), the integration by parts (\ref{oper-9}), and the use of
(\ref{2-appr-8}), (\ref{3-appr-4a}) yield
\begin{eqnarray}
&&\langle(\widetilde{\vu}\cdot\nabla)\,\widetilde{a}_3\rangle=(\overline{\vV}_0\cdot\nabla)\overline{a}_2+
(\overline{\vV}_1\cdot\nabla)\overline{a}_1+\langle(\widetilde{\vxi}\cdot\nabla)(\widetilde{\vxi}\cdot\nabla)\rangle
(\overline{\vV}_0\cdot\nabla)\overline{a}_0-\label{App3-1}\\
&&-\langle(\widetilde{\vxi}\cdot\nabla)(\widetilde{\vxi}_t\cdot\nabla)\rangle\overline{a}_0+
\overline{\mathfrak{X}}\overline{a}_0,\ \text{where}\
\overline{\mathfrak{X}}\equiv\langle(\widetilde{\vxi}\cdot\nabla)(\widetilde{\vu}\cdot\nabla)
\{(\widetilde{\vu}\cdot\nabla)(\widetilde{\vxi}\cdot\nabla)\}^\tau\rangle
\nonumber
\end{eqnarray}
The `Gothic' shorthand operator $\overline{\mathfrak{X}}$ (as well as the operators $\mathfrak{Y}$,
$\mathfrak{\overline{A}}$, $\mathfrak{\overline{B}}$, $\overline{\mathfrak{C}}$, and $\overline{\mathfrak{F}}$ below)
acts on $\overline{a}_0$. The RHS of (\ref{App3-1}) formally contains the fourth, third, second, and first spatial
derivatives of $\overline{a}_0$; however \emph{all the fourth and third derivatives vanish}. In order to prove it we
first rewrite $\overline{\mathfrak{X}}$ as
\begin{eqnarray}
&&\overline{\mathfrak{X}}=\langle(\widetilde{\vxi}\cdot\nabla)(\widetilde{\vu}\cdot\nabla)
\widetilde{\mathfrak{Y}}^\tau\rangle\quad\text{where}\quad
\mathfrak{Y}\equiv(\widetilde{\vu}\cdot\nabla)(\widetilde{\vxi}\cdot\nabla)
\label{App3-2}
\end{eqnarray}
The use of (\ref{oper-9a}) and (\ref{App1-2}), (\ref{App1-1}) transforms (\ref{App3-2}) to
\begin{eqnarray}
&&2\overline{\mathfrak{X}}=-\overline{\mathfrak{A}}+\overline{\mathfrak{B}}+
\frac{1}{2}\langle(\widetilde{\vxi}\cdot\nabla)(\widetilde{\vxi}\cdot\nabla)\rangle(\overline{\vK}\cdot\nabla)
\label{App3-3}\\
&&\overline{\mathfrak{A}}\equiv\langle(\widetilde{\vxi}\cdot\nabla)(\widetilde{\vxi}\cdot\nabla)
(\widetilde{\vu}\cdot\nabla)(\widetilde{\vxi}\cdot\nabla)\rangle,\quad
\overline{\mathfrak{B}}\equiv\langle(\widetilde{\vK}^\tau\cdot\nabla)(\widetilde{\vu}\cdot\nabla)
(\widetilde{\vxi}\cdot\nabla)\rangle
\nonumber
\end{eqnarray}
Let us now simplify $\overline{\mathfrak{A}}$ and $\overline{\mathfrak{B}}$.  For $\overline{\mathfrak{B}}$ we use
(\ref{oper-9a})
\begin{eqnarray}
&&\overline{\mathfrak{B}}\equiv
\langle(\widetilde{\vK}^\tau\cdot\nabla)(\widetilde{\vu}\cdot\nabla)(\widetilde{\vxi}\cdot\nabla)\rangle=-
\langle(\widetilde{\vK}\cdot\nabla)(\widetilde{\vxi}\cdot\nabla)(\widetilde{\vxi}\cdot\nabla)\rangle-
\langle(\widetilde{\vK}^\tau\cdot\nabla)(\widetilde{\vxi}\cdot\nabla)(\widetilde{\vu}\cdot\nabla)\rangle\nonumber
\end{eqnarray}
To change $(\widetilde{\vxi}\cdot\nabla)(\widetilde{\vu}\cdot\nabla)$ into
$(\widetilde{\vu}\cdot\nabla)(\widetilde{\vxi}\cdot\nabla)$ in the last term we use (\ref{App1-2}), (\ref{App1-1}) that
yields:
\begin{eqnarray}
&&\overline{\mathfrak{B}}=-\frac{1}{2}\langle(\widetilde{\vK}\cdot\nabla)(\widetilde{\vxi}\cdot\nabla)
(\widetilde{\vxi}\cdot\nabla)\rangle-\frac{1}{4}\overline{\vkappa}\cdot\nabla,\quad
\vkappa\equiv[\widetilde{\vK}^\tau,\widetilde{\vK}]\label{App3-4}
\end{eqnarray}
The operator $\overline{\mathfrak{A}}$ is simplified by the version of (\ref{oper-9a}) with four multipliers
\begin{eqnarray}
&&\mathfrak{\overline{A}}\equiv\langle(\widetilde{\vxi}\cdot\nabla)(\widetilde{\vxi}\cdot\nabla)
(\widetilde{\vu}\cdot\nabla)(\widetilde{\vxi}\cdot\nabla)\rangle=
-\langle(\widetilde{\vu}\cdot\nabla)(\widetilde{\vxi}\cdot\nabla)(\widetilde{\vxi}\cdot\nabla)
(\widetilde{\vxi}\cdot\nabla)\rangle-\label{App3-5}\\
&&-\langle(\widetilde{\vxi}\cdot\nabla)(\widetilde{\vu}\cdot\nabla)(\widetilde{\vxi}\cdot\nabla)
(\widetilde{\vxi}\cdot\nabla)\rangle-
\langle(\widetilde{\vxi}\cdot\nabla)(\widetilde{\vxi}\cdot\nabla)
(\widetilde{\vxi}\cdot\nabla)(\widetilde{\vu}\cdot\nabla)\rangle
\nonumber
\end{eqnarray}
The multiple use of commutator (\ref{App1-2}), (\ref{App1-1})  allows us to transform the sequence of operators
$(\widetilde{\vxi}\cdot\nabla)$ and $(\widetilde{\vu}\cdot\nabla)$ in each term in the RHS of (\ref{App3-5}) to the
sequence in its LHS. The result is
\begin{eqnarray}
&&\overline{\mathfrak{A}}=-\frac{1}{2}\langle(\vK\cdot\nabla)(\widetilde{\vxi}\cdot\nabla)
(\widetilde{\vxi}\cdot\nabla)\rangle+\frac{1}{4}\overline{\vK''}\cdot\nabla,\quad
\vK''\equiv[\vK',\widetilde{\vxi}]\label{App3-6}
\end{eqnarray}
Now, (\ref{App3-3}), (\ref{App3-4}), and (\ref{App3-6}) yield
\begin{eqnarray}
&&\overline{\mathfrak{X}}=
\frac{1}{4}(\overline{\vK}\cdot\nabla)\langle(\widetilde{\vxi}\cdot\nabla)(\widetilde{\vxi}\cdot\nabla)\rangle+
\frac{1}{4}\langle(\widetilde{\vxi}\cdot\nabla)(\widetilde{\vxi}\cdot\nabla)\rangle(\overline{\vK}\cdot\nabla)-
\frac{1}{8}(\overline{\vkappa}+\overline{\vK''})\cdot\nabla\nonumber
\label{App3-7}
\end{eqnarray}
The substitution of this expression into (\ref{App3-1}), (\ref{4-appr-2})  gives
\begin{eqnarray}
&&\overline{a}_{2s}+(\overline{\vV}_0\cdot\nabla)\overline{a}_2+
(\overline{\vV}_1\cdot\nabla)\overline{a}_1-\frac{1}{8}(\overline{\vkappa}+\overline{\vK''})\cdot\nabla \overline{a}_0
+\frac{1}{4}\overline{\mathfrak{C}}\overline{a}_0-\overline{\mathfrak{F}}\overline{a}_0=0,\label{App3-8}\\
&&\overline{\mathfrak{C}}\equiv(\overline{\vK}\cdot\nabla)\langle(\widetilde{\vxi}\cdot\nabla)(\widetilde{\vxi}\cdot\nabla)
\rangle-\langle(\widetilde{\vxi}\cdot\nabla)(\widetilde{\vxi}\cdot\nabla)\rangle(\overline{\vK}\cdot\nabla),\
\overline{\mathfrak{F}}\equiv\langle(\widetilde{\vxi}\cdot\nabla)(\widetilde{\vxi}_t\cdot\nabla)\rangle
\nonumber
\end{eqnarray}
Additional transformations of the last two operators in (\ref{App3-8}) yield
\begin{eqnarray}
&&\frac{1}{4}\overline{\mathfrak{C}}-\overline{\mathfrak{F}}=
\frac{1}{2}\langle[\widetilde{\vxi},\widetilde{\vxi}_s]\rangle\cdot\nabla
-\frac{1}{2}\langle(\widetilde{\vu}'\cdot\nabla)\widetilde{\vxi}+
(\widetilde{\vxi}\cdot\nabla)\widetilde{\vu}'\rangle\cdot\nabla - \frac{1}{2}\langle
\widetilde{u}'_i\widetilde{\xi}_k+\widetilde{u}'_k\widetilde{\xi}_i\rangle\frac{\partial^2}{\partial x_i\partial x_k}=\nonumber\\
&&=\frac{1}{2}\langle[\widetilde{\vxi},\widetilde{\vxi}_t]\rangle\cdot\nabla-\frac{\partial}{\partial
x_k}\left(\overline{\chi}_{ik}\frac{\partial}{\partial
x_i}\right)+\frac{1}{2}\langle\widetilde{\vxi}\Div\widetilde{\vu}'+\widetilde{\vu}'\Div\widetilde{\vxi}\rangle
\label{Ap-trans}\\
&&\widetilde{\vu}'\equiv\widetilde{\vxi}_s-[\overline{\vV}_0,\widetilde{\vxi}],\quad
\overline{\chi}_{ik}\equiv\frac{1}{2}\langle \widetilde{u}'_i\widetilde{\xi}_k+\widetilde{u}'_k\widetilde{\xi}_i\rangle
 \label{v-prime}
\end{eqnarray}
The  substitution of (\ref{Ap-trans}) into (\ref{App3-8}) leads to the equation for $\overline{a}_2$ (\ref{4.17}) where
the formula (\ref{4.20}) for $\overline{\chi}_{ik}$ is obtained from (\ref{v-prime}) by the use of definition
$\widetilde{\vu}'$.
The tilde-part of (\ref{4-appr-1}) after its tilde-integration gives (\ref{A4.14})
\begin{eqnarray}
&&\widetilde{a}_{4}=-(\widetilde{\vxi}\cdot\nabla)\,\overline{a}_3-
\{(\widetilde{\vu}\cdot\nabla)\,\widetilde{a}_3\}^\tau-\widetilde{a}_{2s}^\tau \label{4-appr-5}
\end{eqnarray}
where $\widetilde{a}_3$ is given by (\ref{3-appr-5}).
Hence, we have solved the equation \eqref{Abasic-2} for the first five approximations and obtained the correspondent required terms in
(\ref{basic-4aa}).

\end{document}